\documentclass[prc,letterpaper,twocolumn,showpacs,showkeys,lengthcheck,tightenlines,
               nofootinbib,preprintnumbers,superscriptaddress]{revtex4-1} 

\usepackage{titlesec}
\usepackage{dcolumn}
\usepackage{bm}
\usepackage{amsmath}
\usepackage{amssymb}
\usepackage{amsfonts}
\usepackage{graphicx}
\usepackage{subfigure}
\usepackage{txfonts}
\usepackage{color}
\usepackage{bbold}

\usepackage{slashed}
\usepackage[svgnames]{xcolor}
\usepackage[english]{babel}
\usepackage{blindtext}
\usepackage{microtype}
\usepackage{tikz}

\usepackage[colorlinks=true,urlcolor=blue]{hyperref}
\usepackage{ifpdf}
\usepackage{float}
\usepackage[utf8]{inputenc}

\def\a{\alpha}
\def\b{\beta}

\def\g{\gamma}

\def\l{\lambda}

\def\o{\omega}

\def\L{\Lambda}

\def\hs{\hspace}

\def\no{\nonumber}

\def\lf{\left}
\def\rg{\right}

\def\lra{\longrightarrow}

\newcommand{\ph}[1]{\phantom{#1}}

\newcommand{\sh}[1]{\slashed{#1}}

\font\bb=bbmss10 scaled 1200
\def\ident{\mbox{\bb 1}}

\graphicspath{{.}{figures/}} 

\titlespacing{\section}{5pt}{12pt plus 4pt minus 2pt}{8pt plus 2pt minus 2pt}
\begin{document}
\title{Flavour dependence of the pion and kaon form factors and parton distribution functions}
\author{Parada~T.~P.~Hutauruk}
\affiliation{CSSM and ARC Centre of Excellence for Particle Physics at the Terascale, \\ 
             Department of Physics, University of Adelaide, Adelaide SA 5005, Australia}

\author{Ian~C.~Clo\"et}
\affiliation{Physics Division, Argonne National Laboratory, Argonne, Illinois 60439, USA}

\author{Anthony~W.~Thomas}
\affiliation{CSSM and ARC Centre of Excellence for Particle Physics at the Terascale, \\ 
             Department of Physics, University of Adelaide, Adelaide SA 5005, Australia}

\begin{abstract}
The separate quark flavour contributions to the pion and kaon valence 
quark distribution functions are studied, along with the corresponding 
electromagnetic form factors in the space-like region. 
The calculations are made using the solution of the Bethe-Salpeter equation for the model of Nambu and Jona-Lasinio with proper-time regularization.  Both the pion and kaon form factors and the valence quark distribution functions reproduce many of the features of the available empirical data. The larger mass if the strange quark naturally explains the empirical  fact that the ratio $u_{K^+}(x)/u_{\pi^+}(x)$ drops below unity at large $x$,  with a value of approximately $M^2_u/M_s^2$ as $x \to 1$. With regard to the elastic form factors we report a large flavour dependence, with the $u$-quark contribution to the kaon form factor being an order of magnitude smaller than that of the $s$-quark at large $Q^2$, which
may be a sensitive measure of confinement effects in QCD. Surprisingly though, the total $K^+$ and $\pi^+$ form factors differ by only 10\%.
\end{abstract}

                             
\maketitle
%===============================================================================
%===============================================================================
\section{\label{intro} INTRODUCTION}

\looseness=-1
In our quest to understand the structure of strongly interacting matter, 
parton distribution functions (PDFs) and electromagnetic form factors are of 
fundamental importance, and provide complementary information. In an infinite 
momentum frame picture the former describe the distribution of longitudinal momentum 
carried by each quark flavour, while the latter are related to their distribution 
transverse to the beam. There have been numerous studies of hadron PDFs and form factors 
within quark models of various degrees of sophistication and success, for example, see Refs.~\cite{Alberg:2011yr,Nam:2012vm,Nguyen:2011jy,Avila:2002xd,Hecht:2000xa,Shigetani:1993dx,Shigetani:1994rp,Davidson:1994uv,Frederico:1994dx,Signal:1989yc,Cloet:2005pp,Bentz:1999gx,Gluck:1991ey,Holt:2010vj,Weigel:1999pc,Berger:1979du,Kusaka:1996vm,Chang:2014gga}
and \cite{Tandy:1997qf,Lemmer:1995eb,Buck:1994hf,Burden:1995ve,Roberts:1994hh,Bijnens:2002hp,Ito:1991pv,Schulze:1994fy,daSilva:2012gf,Dias:2010sg,Wang:2001ne,Zovko:1974gm,Chang:2013nia,Ninomiya:2014kja}, respectively.

\looseness=-1
In this paper we focus on the structure of the pion and kaon, with a particular interest in the effects 
of the larger mass of the strange quark in the kaon. At present, a detailed understanding of 
pion and kaon structure is hampered by the rather small sample of experimental 
data~\cite{Badier:1983mj,Conway:1989fs}. The pion PDF has been measured reasonably well
in the valence region, and it is known that $u_{K^+}(x)$ is somewhat 
softer than $u_{\pi^+}(x)$ in the large-$x$ region. While at the present time one does not 
know the separate flavour contributions to the kaon elastic form factor, it may prove 
possible to measure them in the future, for example, with a parity violating probe.
Further, given the influence of the Drell-Yan-West relation~\cite{Drell:1969km,West:1970av} and its phenomenological 
importance, it is of considerable interest to compare the flavour dependence of 
the large-$x$ PDFs with the corresponding large-$Q^2$ behaviour of the separate flavour 
contributions to the elastic form factor.

We study the structure of the pion and kaon using the Nambu--Jona-Lasinio (NJL)
model with proper-time regularization~\cite{Schwinger:1951nm}
to simulate the effect of quark confinement~\cite{Bentz:2001vc,Ebert:1996vx,Hellstern:1997nv}. The separate contributions of 
each flavour to the pion and kaon elastic form factors are determined with and without the effect of 
vector-meson dressing at the quark-photon vertex. In comparison with existing experimental data 
the model shows excellent agreement. The PDFs are also calculated and the effect 
of the quark masses on the large-$x$ behaviour is explored. We also investigate the effect of 
the spectator quark mass on the PDF for a given quark flavour, finding satisfactory 
agreement with the experimental ratio $u_{K^+}(x)/u_{\pi^+}(x)$. We conclude with a 
discussion of the validity of the Drell-Yan-West relation within this framework.

%===============================================================================
%===============================================================================
\section{NAMBU--JONA-LASINIO MODEL} \label{njl}
The NJL model is a chiral effective theory that mimics many of the 
key features of quantum chromodynamics (QCD) and is therefore a useful tool 
to help understand non-perturbative phenomena in low energy 
QCD~\cite{Klevansky:1992qe,Vogl:1991qt,Vogl:1989ea,Hatsuda:1994pi,Buballa:2003qv}. For example, the NJL model encapsulates
dynamical chiral symmetry breaking, which gives rise to dynamically generated dressed 
quark masses. The NJL model has been successfully used to investigate 
a broad range of phenomena, including hadron properties~\cite{Ishii:1995bu,Mineo:2003vc,Cloet:2005rt,Cloet:2005pp,Cloet:2006bq,Cloet:2007em,Cloet:2014rja,Carrillo-Serrano:2014zta}, heavy ion collisions~\cite{Klahn:2013kga}, neutron stars~\cite{Buballa:2003qv,Lawley:2004bm,Baldo:2006bt}, 
quark fragmentation functions~\cite{Ito:2009zc,Matevosyan:2013aka} and transverse momentum dependent phenomena~\cite{Matevosyan:2011vj}. 

The three-flavour NJL Lagrangian -- containing only four-fermion interactions -- takes the 
form\footnote{In principle the two flavour singlet pieces of the $G_\rho$ term in Eq.~\eqref{NJL lagrangian} can appear in the NJL interaction Lagrangian with separate coupling constants, as they are individually chirally symmetric. Our choice of identical
coupling avoids flavour mixing, giving the flavour content of the $\omega$ meson as $u \bar{u} + d \bar{d}$ and the
$\phi$ meson as $s\bar{s}$.}
\begin{align}
\hs*{-0.8mm}\mathcal{L}_{NJL} = \bar{\psi}(i\slashed{\partial} - \hat{m})\psi 
&+ G_{\pi}\left[(\bar{\psi}\,\lambda_{a}\,\psi)^{2} - (\bar{\psi}\,\lambda_{a}\,\gamma_{5}\,\psi)^{2}\,\right] \no \\
&\hs{0mm}
- G_\rho\left[(\bar{\psi}\,\lambda_{a}\,\gamma^\mu\,\psi)^{2} + (\bar{\psi}\,\lambda_{a}\,\gamma^\mu \gamma_{5}\,\psi)^{2}\right],
\label{NJL lagrangian}
\end{align}
where the quark field has the flavour components $\psi = (u, d, s)$, $\hat{m}={\rm diag} (m_u, m_d, m_s)$ denotes the
current quark mass matrix, and $G_{\pi}, G_\rho$ are four-fermion coupling constants. A sum over $a = 0,\ldots,8$ is implied in 
Eq.~\eqref{NJL lagrangian}, where $\lambda_1,\dots,\lambda_8$ are the Gell-Mann matrices in flavour space
and $\lambda_0 \equiv \sqrt{\frac{2}{3}}\,\ident$. The elementary quark-antiquark interaction kernel derived from 
Eq.~\eqref{NJL lagrangian} takes the form 
\begin{align}
\label{eq:kernel}
&\mathcal{K}_{\alpha\beta,\gamma\delta} = \sum_{\Omega}\, \mathcal{K}_{\Omega}\, \Omega_{\gamma \delta}\, \bar{\Omega}_{\alpha \beta} \no \allowdisplaybreaks \\
&\hs{2mm}
= 2i\,G_\pi \lf[(\l_a)_{\gamma \delta}\,(\l_a)_{\alpha \beta} + (\l_a\,\gamma_5)_{\gamma \delta}\,(\l_a\,\gamma_5)_{\alpha \beta}\rg] \no \allowdisplaybreaks \\
&\hs{2mm}
- 2i\,G_\rho\lf[(\l_a\,\g^\mu)_{\gamma \delta}\,(\l_a\,\g_\mu)_{\alpha \beta} + (\l_a\,\g^\mu\gamma_5)_{\gamma \delta}\, (\l_a\,\g_\mu\gamma_5)_{\alpha \beta}\rg],
\end{align}
where the indices represent Dirac, colour and flavour. In this work we assume that $m_u = m_d = m$, and with the Lagrangian 
of Eq.~\eqref{NJL lagrangian} the $\rho$ and $\omega$ mesons are therefore mass degenerate, differing only in their flavour structure.

The NJL model is non-renormalizable and hence a regularization scheme must be used to control divergences. Here the  
proper-time scheme is chosen, because it simulates aspects of quark confinement by eliminating 
on-shell quark propagation, while maintaining the symmetries of the theory, such as the Poincar\'e and chiral symmetries. 
As a result it has been widely used~\cite{Ebert:1996vx,Hellstern:1997nv,Horikawa:2005dh,Weiss:1993kv,Ripka:1996fw,Broniowski:1995yq,Bijnens:1995ww,Ebert:1994tv,Cloet:2014rja}. 
Formally the proper-time regularization scheme is defined by
\begin{align}
\label{eq:propertimeregularization}
\frac{1}{X^n} &= \frac{1}{(n-1)!} \int_0^\infty d\tau\, \tau^{n-1} e^{-\tau X} \no \\
&\hs{25mm}
\longrightarrow\ \frac{1}{(n-1)!} 
\int_{1/\Lambda_{\rm UV}^2}^{1/\Lambda_{\rm IR}^2} 
d\tau\, \tau^{n-1} e^{-\tau X},
\end{align} 
where $X^n$ is obtained by first introducing Feynman parametrization and then
performing a Wick rotation of the loop momenta to Euclidean space. Only 
the ultraviolet cutoff, $\Lambda_{\rm UV}$, is needed to render the 
theory finite. However, in bound states of quarks we also include 
the infrared cutoff, $\Lambda_{\rm IR}$, which eliminates unphysical thresholds 
for the decay of hadrons into quarks, therefore implementing quark confinement
in the NJL model.

%===============================================================================
\begin{figure}[tbp]
\centering\includegraphics[width=\columnwidth]{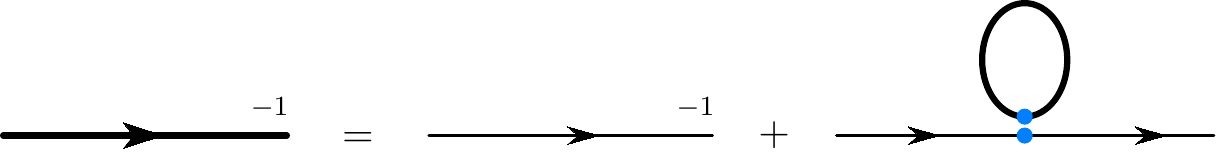}
\caption{ (Colour online) The NJL gap equation in the Hartree-Fock approximation. The thin line is the bare quark 
propagator, $S_0^{-1}(k) = \sh{k} - m + i\varepsilon$, 
whereas the thick line is the dressed quark propagator $S(k)$. 
The $\bar{q}q$ interaction kernel is given by Eq.~\eqref{eq:kernel}.}
\label{fig:Gape1}
\end{figure}
%===============================================================================

The standard NJL gap equation, illustrated in Fig.~\ref{fig:Gape1}, provides the dressed quark propagator.
The general solution of this gap equation has the form $S_q^{-1}(p) = \sh{p} - M_q + i\varepsilon$,
where the dressed quark mass for each quark flavour $q = u,d,s$ satisfies
\begin{align}
\label{eq:gapequation}
M_q &= m_q - 4\,G_\pi\, \langle \bar{q}q \rangle
= m_q + 12 i\,G_\pi\!\!\int\!\frac{d^4k}{(2\pi)^4}\, \mathrm{Tr}_{D}[S_q(k)].
\end{align}
The quark condensate is denoted by $\langle \bar{q}q \rangle$ and $m_q$ is the current mass 
for each quark flavour. Introducing the proper-time regularization scheme gives
\begin{align}
\label{eq:gapmass}
M_q &= m_q + \frac{3\,M_q\,G_\pi}{\pi^2} 
\int d\tau \
\frac{1}{\tau^2}\ e^{-\tau\,M_q^2},
\end{align}
where here, and in the following, we drop the proper-time regularization parameters to aid readability.
In the chiral limit ($\hat{m}=0$) the NJL Lagrangian respects the chiral $S\!U(3)_L \otimes S\!U(3)_R$ symmetry,
however a non-trivial solution ($M_q \neq 0$) to Eq.~\eqref{eq:gapequation} exists provided $G_\pi > G_{\text{critical}}$,
which is a signature for dynamical chiral symmetry breaking (DCSB).

\begin{figure}[tbp]
\centering\includegraphics[width=\columnwidth]{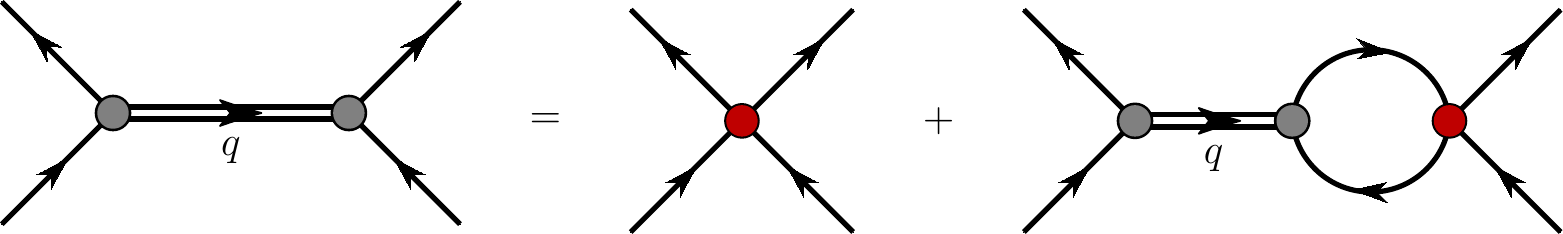}
\caption{(Colour online) The Bethe-Salpeter equation illustrated here for quark and antiquark scattering.}
\label{fig:bse1} 
\end{figure}

The mesons considered here -- $\pi$, $K$, $\rho$, $\omega$ and $\phi$ -- are realized in the NJL model as quark-antiquark bound states whose properties are governed by the Bethe-Saltpeter equation (BSE) illustrated in Fig.~\ref{fig:bse1}.
This BSE takes the form
\begin{align}
\label{eq:bsemesons}
T(q) &= 
\mathcal{K} + \int \frac{d^4k}{(2\pi)^4}\ \mathcal{K} \ S(k+q)\ T(q) \ S(k),
\end{align}  
where $q$ is the total 4-momentum of the two-body system and the 
Dirac, colour and flavour indices have been omitted.

%===============================================================================
%===============================================================================

The solution to the Bethe-Salpeter equation in the $\a = \pi,\,K$ and $\b = \rho,\,\omega,\,\phi$ channels,
 are respectively
\begin{align}
T_\a(q)_{a b, c d} &= 
\left[ \gamma_5\,\l_\a\right]_{a b} \tau_\a (q) \left[ \gamma_5\,\l_\a^\dagger \right]_{c d}, \\
T_\b(q)_{a b, c d} &= 
\left[ \gamma_\mu\,\l_\b \right]_{a b} \tau_\b^{\mu\nu}(q) 
\left[ \gamma_\nu\,\l_\b^\dagger \right]_{c d},
\end{align}
where $\l_\a,\,\l_\b$ are the appropriate flavour matrices, for example,
$\l_{\pi^0} = \lambda_3$,  $\l_{\pi^\pm} = \frac{1}{\sqrt{2}} (\lambda_1 \pm i \lambda_2)$ and
$\l_{K^\pm} = \frac{1}{\sqrt{2}} (\lambda_4 \pm i \lambda_5)$. The reduced 
$t$-matrices in these channels take the form
\begin{align}
\label{eq:tmatrix}
\tau_\a (q) &= \frac{-2i\,G_\pi}{1 + 2\,G_\pi\,\Pi_\a (q^2)}, \\
\tau_\b^{\mu \nu} (q) &= \frac{-2i\,G_\rho}{1 + 2\,G_\rho\,\Pi_\b (q^2)} \left(g^{\mu \nu} + 2\,G_{\rho}\,\Pi_\b(q^2)\, \frac{q^{\mu} 
q^{\nu}}{q^2} \right),
\end{align}
where the bubble diagrams appearing read:
\begin{align}
\label{eq:bubblegraphtot}
\Pi_{\pi}(q^2) &= 6i \int \frac{d^4k}{(2\pi)^4}\ \mathrm{Tr}_D\left[\gamma_5\,S_{\!\ell}(k) \gamma_5\,S_{\!\ell}(k+q) \right], \\
\Pi_{K} (q^2) &= 6i \int \frac{d^4k}{(2\pi)^4}\ \mathrm{Tr}_D\left[\gamma_5\,S_{\!\ell}(k) \gamma_5\,S_{\!s}(k+q) \right], \\
\Pi_{v}^{aa} (q^2)\, P_T^{\mu\nu} &= 6i\int \frac{d^4 k}{(2\pi)^4}\ \mathrm{Tr}_D \left[\gamma^\mu S_{\!a} (k) \gamma^\nu S_{\!a}(k+q) \right],
\end{align}
where $\Pi_\rho = \Pi_\o = \Pi_{v}^{\ell\ell}$, $\Pi_\phi = \Pi_{v}^{ss}$ and $\ell \equiv u,\,d$. The trace is over Dirac indices only and
$P_T^{\mu\nu} = g^{\mu\nu} - q^\mu q^\nu/q^2$.

The meson masses are defined by the pole in the corresponding $t$-matrix, and 
therefore the pion mass, for example, is determined by the pole condition:
\begin{align}
1 + 2\, G_\pi\, \Pi_{\pi} (k^2 = m_{\pi}^2) &= 0,
\end{align}  
where analogous results determine $m_K$, $m_\rho$, $m_\omega$ and $m_\phi$. In fact, because of DCSB the pion 
and kaon masses are given by the simple expressions:
\begin{align}
m_{\pi}^2 &= \left[ \frac{m}{M_\ell} \right] 
\frac{2}{G_\pi\, \mathcal{I}_{\!\ell\ell}(m_\pi^2)}, \\
m_{K}^2 &= \left[ \frac{m_s}{M_s} + \frac{m}{M_\ell} \right] 
\frac{1}{G_\pi\, \mathcal{I}_{\!\ell s}(m_K^2)} + (M_s-M_\ell)^2,
\end{align} 
where
\begin{align}
\mathcal{I}_{\!ab}(k^2) &= \frac{3}{\pi^2} \int_0^1 dx \int
 \frac{d\tau}{\tau}\
e^{-\tau\lf[x(x-1)\,k^2 + x\,M_b^2 + (1-x)\,M_a^2\rg]}.
\end{align} 
This demonstrates the Goldstone boson nature of the pion and kaon in the chiral limit.
The residue at a pole in the $\bar{q}q$ $t$-matrices defines the effective meson-quark-quark coupling 
constant, and for the various mesons we obtain
\begin{align}
\label{eq:couplinconstant}
Z_{\a}^{-1} &= -\left.\frac{\partial\, \Pi_\a(q^2)}{\partial q^2} \right|_{q^2 = m_\a^2}, \quad \a = \pi,\,K, \,\rho,\,\omega,\,\phi. 
\end{align}

The parameters of our NJL model are therefore: the couplings in the NJL Lagrangian
$G_\pi$ and $G_\rho$; the regularization parameters $\Lambda_{\rm{IR}}$ and 
$\Lambda_{\rm UV}$; and the $u/d$ and $s$ dressed quark masses (or alternatively their current
quark masses). In QCD the confinement scale is set by $\L_{\rm QCD}$ and therefore we
fix $\Lambda_{\rm IR} = 240\,$MeV and choose the dressed light quark mass as 
$M = 400\,$MeV. The remaining parameters are then fit to the physical pion 
($m_\pi = 140\,$MeV), kaon ($m_K = 495\,$MeV) and rho ($m_\rho = 770\,$MeV) masses, 
together with the pion decay constant ($f_\pi = 93\,$MeV). This gives
$G_\pi = 19.04\,$GeV$^{-2}$, $G_\rho = 11.04\,$GeV$^{-2}$, 
$\Lambda_{\rm UV} = 645\,$MeV and $M_s = 611\,$MeV.

Elementary results in this NJL model are presented in Tab.~\ref{tab:parameters}. A focus 
herein is the effect of explicit chiral symmetry and flavour symmetry violation. As a starting point 
we can consider the Goldberger--Treiman relation at the quark level, and the
Gell-Mann--Oakes--Renner relation. For the pion these read
\begin{align}
f_\pi\sqrt{Z_\pi} &= \frac{1}{2}\lf(M_u + M_d\rg),  \\ 
f_\pi^2\,m_\pi^2 &= -\frac{1}{2}\lf(m_u + m_d\rg)\bigl<\bar{u}u + \bar{d}d\bigr>,
\end{align}
and in the chiral limit these relations are satisfied exactly. With the parameters
above we find violation at the 1\% level for the pion. However, for the analogous relations for
the kaon we find violations at the 20-25\% level, which is sizeable, but much less than 
what may be expected from the current quark mass ratio $2\,m_s / (m_u + m_d) = 27.5\pm 1.0$~\cite{Beringer:1900zz,Durr:2010vn}.

%===============================================================================
\begin{table}[tbp]
\addtolength{\tabcolsep}{2.9pt}
\addtolength{\extrarowheight}{2.2pt}
\begin{tabular}{cccccccccc}
\hline\hline
$Z_\pi$ & $Z_K$ & $Z_\rho$ & $Z_\o$ & $Z_\phi$ & $f_K$  & $\lf<\bar{u}u\rg>^{1/3}$  & $\lf<\bar{s}s\rg>^{1/3}$ \\[0.2em]
\hline
17.85  & 20.89 & 8.44 & 8.44 & 13.02 & 0.097 & $-$0.171 & $-$0.150 \\
\hline\hline
\end{tabular}
\caption{Results for the meson-quark-quark coupling constants, kaon leptonic decay 
constant and the quark condensates. All dimensioned quantities are in units of GeV.}
\label{tab:parameters}
\end{table}
%===============================================================================

%===============================================================================
%===============================================================================
\section{ELASTIC FORM FACTORS} \label{formfactor}
To determine the electromagnetic current of the pion or kaon we couple
the electromagnetic field to the quark fields via minimal substitution:
 $i\sh{\partial} \rightarrow i \sh{\partial} - \hat{Q}\, A_\mu\,\gamma^\mu$, 
where $A_\mu$ is electromagnetic potential, $e$ is the positron charge
and $\hat{Q} = \text{diag}\lf[e_u,\,e_d,\,e_s\rg] = \frac{e}{2} (\lambda_3 + \frac{1}{\sqrt{3}} \lambda_8)$ is the 
quark charge operator, where $e_q$ are the quark charges.
The matrix element of the electromagnetic current for a pseudoscalar meson reads
\begin{align}
\label{eq:formfactor1}
J_\a^{\mu} (p',p) &= \lf(p'^\mu + p^\mu\rg) F_\a (Q^2), \quad \a = \pi,\,K,
\end{align}
where $p$ and $p'$ denote the initial and final four momenta of the pseudoscalar meson, 
$q^2 =(p'-p)^2 \equiv -Q^2$ and $F_\a (Q^2)$ is the pion or kaon form factor.

\begin{figure}[tbp]
\centering\includegraphics[width=\columnwidth]{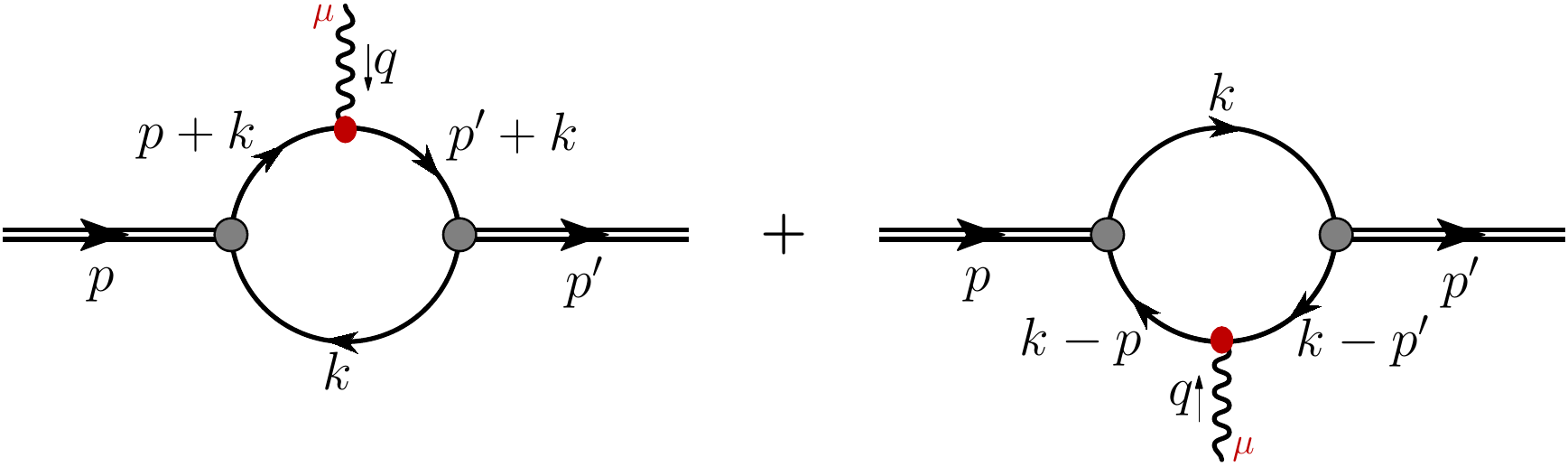}
\caption{(Colour online) Diagrammatic representation of the 
electromagnetic current of the pion or kaon.}
\label{fig:emvertex1}
\end{figure}

The pseudoscalar meson form factors in the NJL model are given by the sum of the 
two Feynman diagrams depicted in Fig~\ref{fig:emvertex1}, which are respectively given by
\begin{align}
\label{eq:j1}
j^{\mu}_{1,\a}\lf(p',p\rg) &= i\,Z_\a \int \frac{d^4k}{(2\pi)^4} \no \\
&\hs{-1mm}
\mathrm{Tr}\lf[\g_5\,\l_\a^\dagger\,S(p'+k)\,\hat{Q}\,\g^\mu\,S(p+k)\,\g_5\,\l_\a\,S(k)\rg], \\
\label{eq:j2}
j^{\mu}_{2,\a}\lf(p',p\rg) &= i\,Z_\a \int \frac{d^4k}{(2\pi)^4} \no \\
&\hs{-1mm}
\mathrm{Tr}\lf[\g_5\,\l_\a\,S(k-p)\,\hat{Q}\,\g^\mu\,S(k-p')\,\g_5\,\l_\a^\dagger\,S(k)\rg],
\end{align}
where the trace is over Dirac, colour and flavour indices. The index $\a$ labels the state and 
$\l_\a$ are the corresponding flavour matrices. In flavour space the quark propagator reads $S(p) = \text{diag}[S_u(p),\,S_d(p),\,S_s(p)]$.

%===============================================================================
%===============================================================================

We will focus on the quark sector and total form factors for $\pi^+$, $K^+$ 
and $K^0$, where for the form factors we find
\begin{align}
\label{eq:barepi}
F_{\pi^{+}}^{\text{(bare)}}(Q^2) &= \lf(e_u - e_d\rg) f^{\ell\ell}_\pi(Q^2), \\
F_{K^{+}}^{\text{(bare)}}(Q^2) &= e_u\,f^{\ell s}_K(Q^2) - e_s\,f^{s\ell}_K(Q^2),\\
\label{eq:bareK0}
F_{K^0}^{\text{(bare)}}(Q^2) &= e_d\,f^{\ell s}_K(Q^2) - e_s\,f^{s\ell}_K(Q^2).
\end{align}
The first superscript on the body form factors, $f_\a^{ab}(Q^2)$, indicates the struck quark 
and the second the spectator, where
\begin{align}
&f^{ab}_\a(Q^2) = \frac{3\,Z_\a}{4\,\pi^2} \int_0^1 dx \int \frac{d\tau}{\tau}\ e^{-\tau\lf[M_a^2 + x(1-x)\,Q^2\rg]} \no \\
&+ \frac{3\,Z_\a}{4\,\pi^2}  \int_0^1\! dx\! \int_0^{1-x}\! dz\! \int\! d\tau\,
\Bigl[(x+z)\,m_\a^2 + (M_a - M_b)^2(x+z) \nonumber \\
&
+ 2\,M_b\lf(M_a - M_b\rg)\Bigr]\,
e^{-\tau\lf[(x+z)(x+z-1)\,m_\a^2 + (x+z)\,M_a^2 + (1-x-z)\,M_b^2 + x\,z\,Q^2\rg]}.
\end{align}
These results are denoted as ``bare'' because the quark-photon vertex is the 
elementary result, that is,  $\L_{\g q}^{\mu\text{(bare)}} = \hat{Q}\,\g^\mu$. 
Importantly, these expressions satisfy charge conservation exactly.

The quark-sector form factors for a hadron $\a$ are defined by
\begin{align}
F_\a(Q^2) = e_u\,F_\a^u(Q^2) + e_d\,F_\a^d(Q^2) + e_s\,F_\a^s(Q^2) + \ldots
\label{eq:quarksector}
\end{align}
Therefore the ``bare'' pseudoscalar meson quark-sector form factors are easily read off from
Eqs.~\eqref{eq:barepi}-\eqref{eq:bareK0}.

%===============================================================================
\begin{figure}[tbp]
\centering\includegraphics[width=\columnwidth]{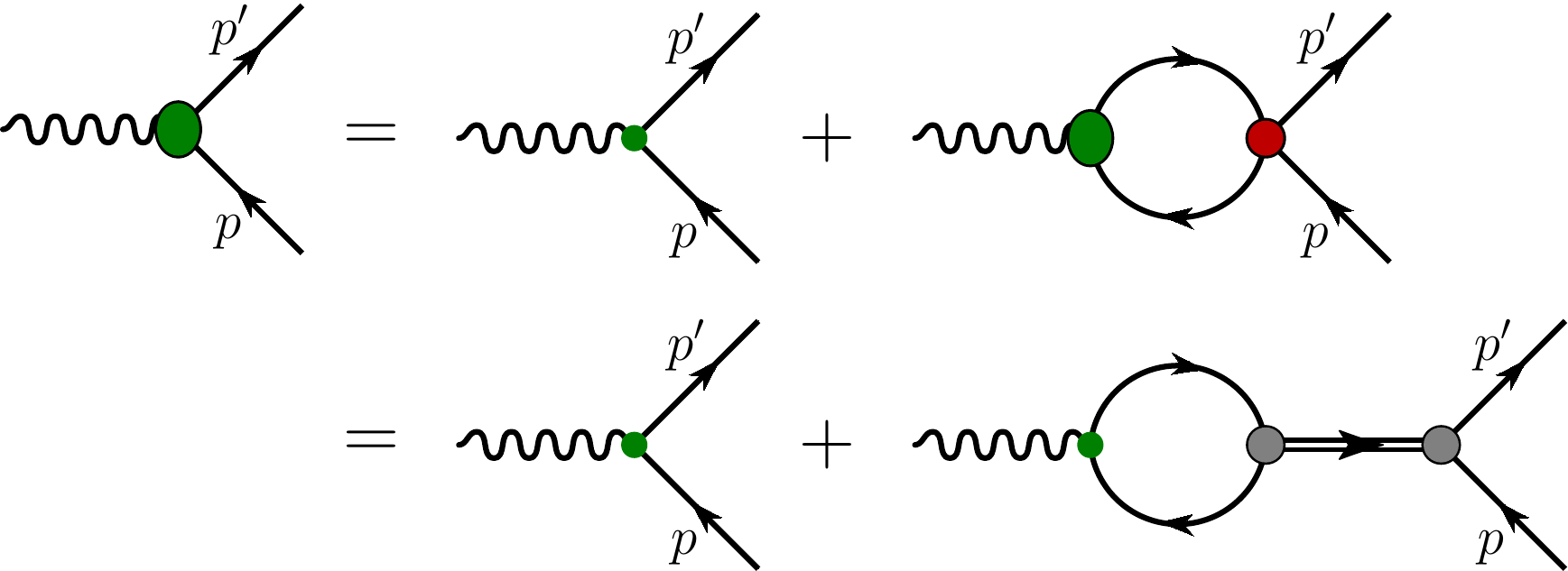}
\caption{(Colour online) Illustration of the inhomogeneous BSE which gives the dressed quark-photon vertex.  
The large shaded oval represents the solution of the inhomogeneous BSE, the small dot is the inhomogeneous driving term ($\hat{Q}\,\g^\mu$) and the double-dots represent the $q\bar{q}$ interaction kernel given in Eq.~\eqref{eq:kernel}.}
\label{fig:vectormesons}
\end{figure}
%===============================================================================

\begin{figure}[tbp]
\centering\includegraphics[width=\columnwidth]{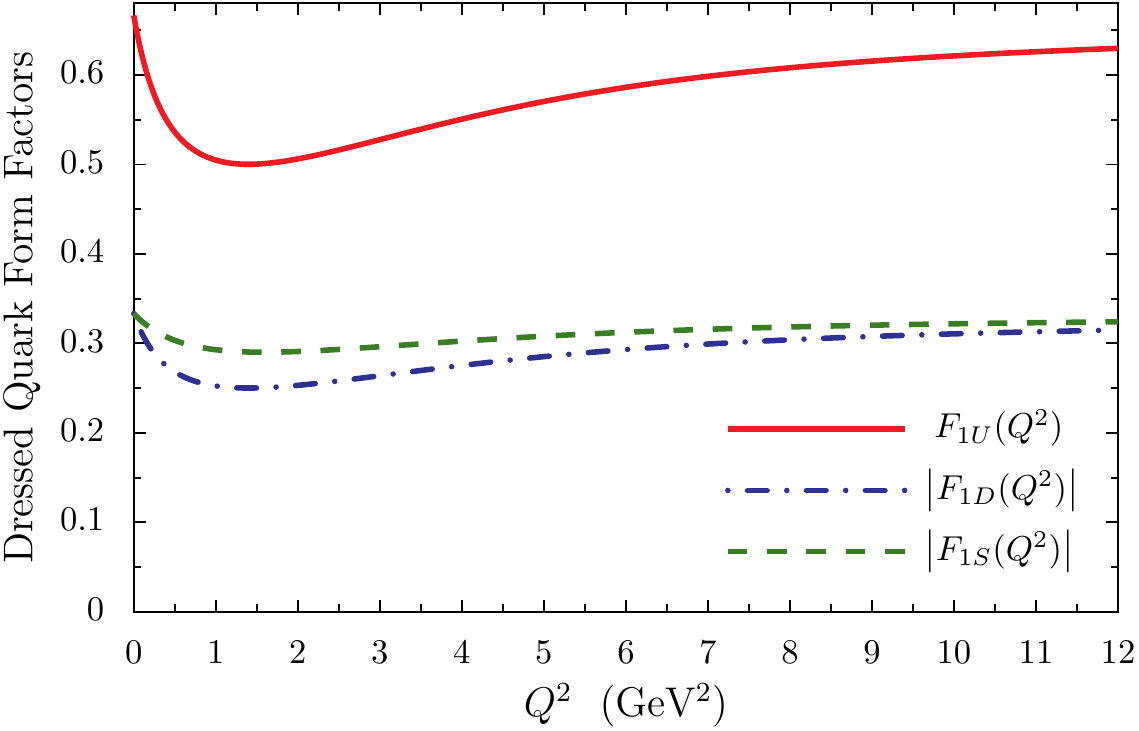}
\caption{(Colour online) The dressed quark form factors obtained as solutions to the inhomogenerous
Bethe-Salpeter equation.}
 \label{fig:dressedformfactors}
\end{figure}

In general the quark-photon vertex is not elementary ($\hat{Q}\,\g^\mu$) but is 
instead dressed, with this dressing given by the inhomogenerous Bethe-Salpeter 
equation, which is illustrated in Fig.~\ref{fig:vectormesons}. With the NJL kernel
of Eq.~\eqref{eq:kernel}, the general solution for the dressed quark-photon vertex
for a quark of flavour $q$, has the form
\begin{align}
\label{eq:quarkvertex}
\hs{-1mm}\L^\mu_{\g\,Q}(p',p) &= e_q\,\g^\mu + \lf(\g^\mu - \frac{q^\mu\sh{q}}{q^2}\rg)F_{Q}(Q^2)
\to \g^\mu\,F_{1Q}(Q^2),
\end{align}
where the final result is used because the $q^\mu\sh{q}/q^2$ term cannot contribute
to a hadron electromagnetic current because of current conservation. Note, the
result after the equality in Eq.~\eqref{eq:quarkvertex} clearly satisfies the Ward-Takahashi identity:
\begin{align}
q_\mu\,\L^\mu_{\g\,Q}(p',p) = e_q\lf[S_q^{-1}(p') - S_q^{-1}(p)\rg].
\end{align}
For the dressed $u$, $d$ and $s$ quarks we find
\begin{align}
\label{eq:f1U}
F_{1U/D}(Q^2) &= e_{u/d}\ \frac{1}{1 + 2\,G_\rho\,\Pi_{v}^{\ell\ell}(Q^2)}, \\
%
%\label{eq:f1D}
%F_{1D}(Q^2) &= e_d\ \frac{1}{1 + 2\,G_\rho\,\Pi_{v}^{\ell\ell}(Q^2)}, \\
%
\label{eq:f1S}
F_{1S}(Q^2) &= e_s\ \frac{1}{1 + 2\,G_\rho\,\Pi_{v}^{ss}(Q^2)},
\end{align}
where the explicit form of the bubble diagram is
\begin{align}
\hs*{-1mm}\Pi^{qq}_{v} (Q^2) &= \frac{3\,Q^2}{\pi^2} \int_0^1\!\! dx \int\!\! \frac{d\tau}{\tau}\
x\lf(1-x\rg)\, e^{-\tau\lf[M_q^2 + x\lf(1-x\rg)Q^2\rg]}.
\end{align}
Therefore, with the NJL Lagrangian of Eq.~\eqref{NJL lagrangian} there is no flavour
mixing in the dressed quark form factors, in analogy with the dressed quark masses.
The dressed quark form factors are illustrated in Fig.~\ref{fig:dressedformfactors}.
In the limit $Q^2 \to \infty$ these form factors reduce to the elementary 
quark charges, as expected because of asymptotic freedom in QCD. For small
$Q^2$ these results are similar to expectations form vector meson dominance,
where the dressed $u$ and $d$ quarks are dressed by $\rho$ and $\omega$ mesons and
the dressed $s$ quark by the $\phi$ meson. Note, the denominators in 
Eqs.~\eqref{eq:f1U} and \eqref{eq:f1S} are the same as the pole condition 
obtained by solving the Bethe-Salpeter equation in the $\rho$, $\omega$ or $\phi$ channels. Therefore,
the dressed $u$ and $d$ quark form factors have poles at $Q^2 = -m_\rho^2 = -m_\o^2$, and the 
dressed $s$ quark form factor has a pole at $Q^2 = -m_\phi^2$.

%===============================================================================
%===============================================================================

The complete results for the pseudoscalar meson form factors -- with a dressed quark-photon vertex -- read
\begin{align}
\label{eq:fullpi}
F_{\pi^{+}}(Q^2) &= \lf[F_{1U}(Q^2) - F_{1D}(Q^2)\rg] f^{\ell\ell}_\pi(Q^2),  \\
F_{K^{+}}(Q^2) &= F_{1U}(Q^2)\,f^{\ell s}_K(Q^2)
- F_{1S}(Q^2)\,f^{s\ell}_K(Q^2), \\
\label{eq:fullK0}
F_{K^0}(Q^2) &= F_{1D}(Q^2)\,f^{\ell s}_K(Q^2)
 - F_{1S}(Q^2)\,f^{s\ell}_K(Q^2),
\end{align}
where the quark-sector form factors are easily obtained by noting Eq.~\eqref{eq:quarksector} 
and the results in Eqs.~\eqref{eq:f1U}-\eqref{eq:f1S}.

%===============================================================================
%===============================================================================
\section{VALENCE QUARK DISTRIBUTIONS OF THE KAON\label{structurefunction}} 
The twist-2 quark distributions in a hadron $\a$ are defined by
\begin{align}
\label{eq:valence1}
q_\a(x) &= p^{+} \int \frac{d\xi^{-}}{2\pi}\ e^{ix\,p^{+}\,\xi^{-}}\
\langle \a\,| \bar{\psi}_{q}(0) \gamma^{+} \psi_q (\xi^{-})|\,\a \rangle_{c},
\end{align}
where $q$ is the quark flavour, $c$ denotes a connected matrix element and $x = \frac{k^+}{p^+}$ is the Bjorken 
scaling variable, where $p^+$ is the plus-component of the hadron momentum 
and $k^+$ is the plus-component of the struck quark momentum.
Note, in the NJL model the gluons are ``integrated out'' and therefore the gauge-link 
which should appear in Eq.~\eqref{eq:valence1} is unity.

%===============================================================================
\begin{figure}[bp]
\centering\includegraphics[width=\columnwidth]{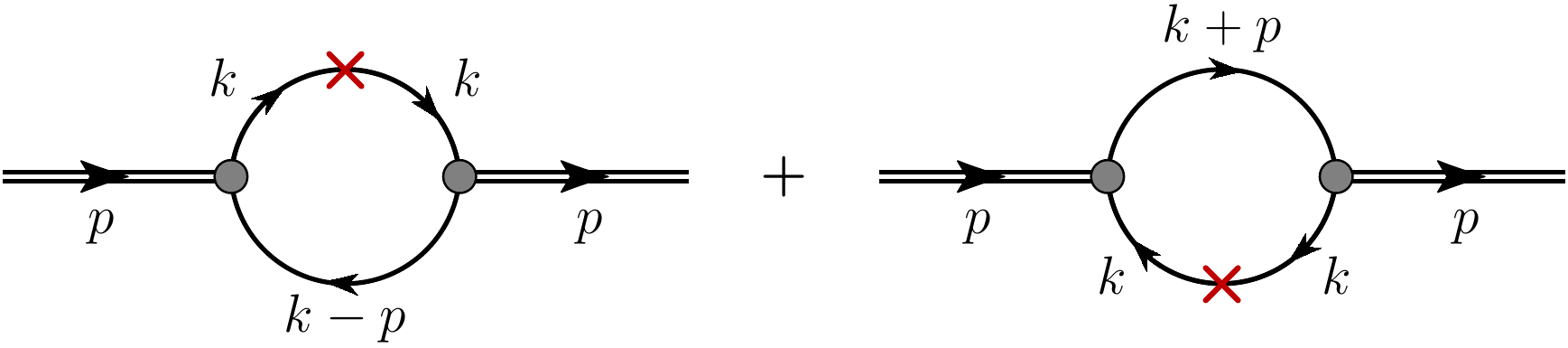}
\caption{\label{fig:strucfun1} (Colour online) Feynman diagrams for the 
valence quark distributions in the pion or kaon.
The red cross is the operator insertion 
$\g^+ \delta \lf(p^+x - k^+\rg)\hat{P}_q$, 
where $\hat{P}_q$ is the projection operator for quarks of flavour $q$.}
\end{figure}
%===============================================================================

From Eq.~\eqref{eq:valence1} one may readily show that the valence quark distribution functions of the pion or kaon are given by
the two Feynman diagrams in Fig.~\ref{fig:strucfun1}, where the operator insertion
is given by $\g^+\delta\lf(p^+x - k^+\rg)\hat{P}_q$ and $\hat{P}_q$ is the projection operator 
for quarks of flavour $q$:
\begin{align}
\hat{P}_{u/d} &= \frac{1}{2}\lf(\frac{2}{3}\,\ident \pm \l_3 + \frac{1}{\sqrt{3}}\,\l_8\rg), &
\hat{P}_s &= \frac{1}{3}\,\ident - \frac{1}{\sqrt{3}}\,\l_8.
\end{align}
Using the relation $\bar{q}(x) = -q(-x)$ the valence quark and anti-quark distributions 
in the pion or kaon are given by
\begin{align}
\label{eq:valence3}
q_\a(x) &= i\,Z_\a \int \frac{d^4k}{(2\pi)^4}\ \delta\lf(p^+x - k^+\rg) \no \\
&\hs{9mm}
\times \mathrm{Tr}\lf[\gamma_5\l_\a^\dagger\,S(k)\,\g^+\hat{P}_q\,S(k)\,\gamma_5\l_\a\,S(k-p)\rg], \\
\bar{q}_\a(x) &= -i\,Z_\a \int \frac{d^4k}{(2\pi)^4}\ \delta\lf(p^+x + k^+\rg) \no \\
&\hs{9mm}
\times \mathrm{Tr}\lf[\gamma_5\l_\a\,S(k)\,\g^+\hat{P}_q\,S(k)\,\gamma_5\l_\a^\dagger\,S(k+p)\rg].
\end{align}   

To evaluate these expressions we first take the moments:
\begin{align}
\label{eq:valence2}
\mathcal{A}_n &= \int_0^1 dx\, x^{n-1}\, q(x),
\end{align}
where $n = 1,\,2,\ldots$ is an integer. Using the Ward-like identity 
$S(k)\,\g^+\,S(k) = -\partial\,S(k)/\partial k_+$ and introducing the Feynman
parametrization, the quark and anti-quark distributions can then be straightforwardly obtained.
For the valence quark and anti-quark distributions of the $K^+$ we find:
\begin{align}
\label{eq:valence5}
q_{K^+}(x)  &= \frac{3\,Z_K}{4\pi^2}  \int d\tau\
e^{-\tau\lf[x(x - 1)\,m_K^2 + x\,M_s^2 + (1-x)\,M_\ell^2\rg]} \nonumber \\
&\hs{14.5mm}
\times \left[\frac{1}{\tau} + x(1 - x)\left[m_K^2 - (M_\ell - M_s)^2\right] \right], \\ 
\bar{q}_{K^+}(x)  &= \frac{3\,Z_K}{4\pi^2}  \int d\tau\
e^{-\tau\lf[x(x - 1)\,m_K^2 + x\,M_\ell^2 +  (1-x)\,M_s^2\rg]} \nonumber \\
&\hs{14.5mm}
\times \left[\frac{1}{\tau} + x(1 - x)\left[m_K^2 - (M_\ell - M_s)^2\right]\right].
\end{align}
Results for the $\pi^+$ are obtained by $M_s \to M_\ell$ and $Z_K \to Z_\pi$, giving the result
$u_{\pi^+}(x) = \bar{d}_{\pi^+}(x)$. The quark distributions for the other 
pseudoscalar mesons can be obtained using flavour symmetries.

The quark distributions satisfy the baryon number and momentum sum rules, which
for the $K^+$ read:
\begin{align}
\label{eq:valence6}
\hs*{-1mm}\int_0^1\!\! dx\lf[u_{K^{+}}(x) - \bar{u}_{K^+}(x)\rg] =
\int_0^1\!\! dx\lf[\bar{s}_{K^{+}}(x) - s_{K^+}(x)\rg] = 1,
\end{align}
for the number sum rule and at the model scale the momentum sum rule is give by
\begin{align}
\label{eq:valence7}
&\int_0^1 dx\ x\lf[u_{K^+}(x) + \bar{u}_{K^+}(x) + s_{K^+}(x) + \bar{s}_{K^+}(x)\rg] = 1. 
\end{align}
Analogous results hold for the remaining kaons and the pions.

%===============================================================================
%===============================================================================
\section{ELASTIC FORM FACTORS RESULTS\label{result}} 
\begin{figure}[tbp]
\centering\includegraphics[width=\columnwidth]{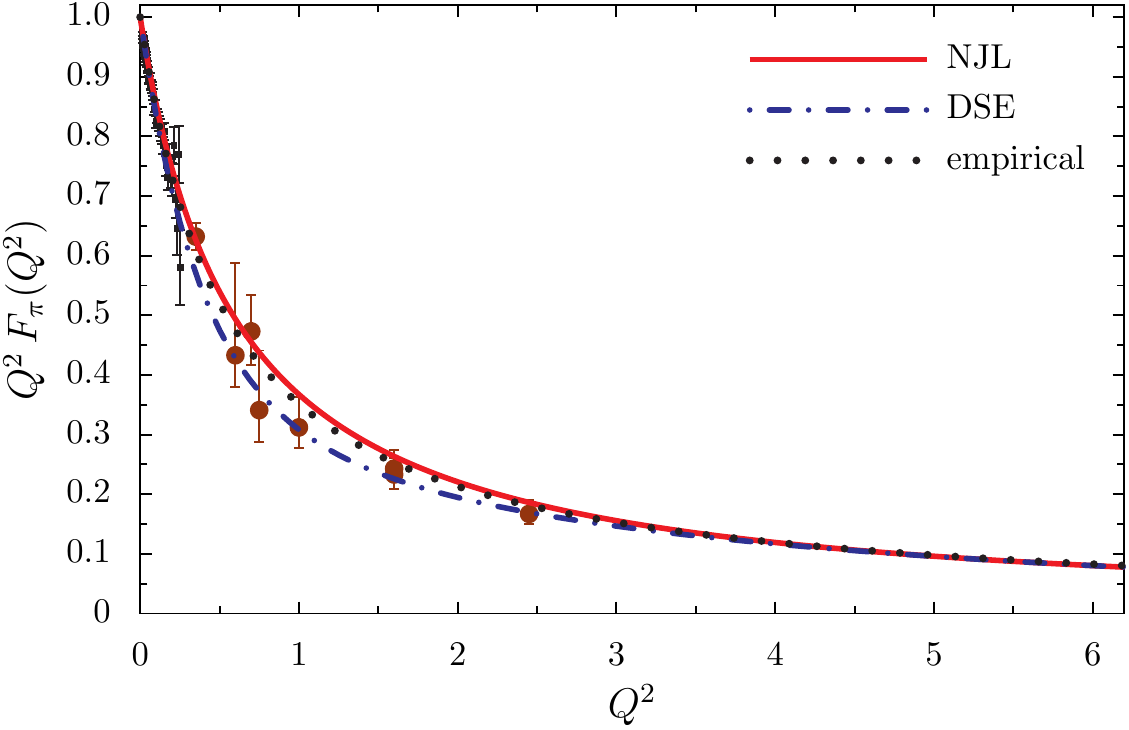}
\caption{(Colour online)
Our results for the pion form factor are given as the solid line and a comparison 
is made to the DSE results of Ref.~\cite{Chang:2013nia}. The empirical result (dotted line) has the 
form $F_\pi(Q^2) = [1 + Q^2/\L_\pi^2]^{-1}$, where the mass parameter is chosen to reproduce
empirical radius found in Ref.~\cite{Amendolia:1984nz}, giving $\L_\pi^2 = 0.54\,$GeV$^2$. The experimental data is from 
Refs.~\cite{Amendolia:1984nz,Amendolia:1986wj,Horn:2006tm,Tadevosyan:2007yd,Huber:2008id,Blok:2008jy}.}
\label{fig:pionff}
\end{figure}

\begin{figure}[tbp]
\centering\includegraphics[width=\columnwidth]{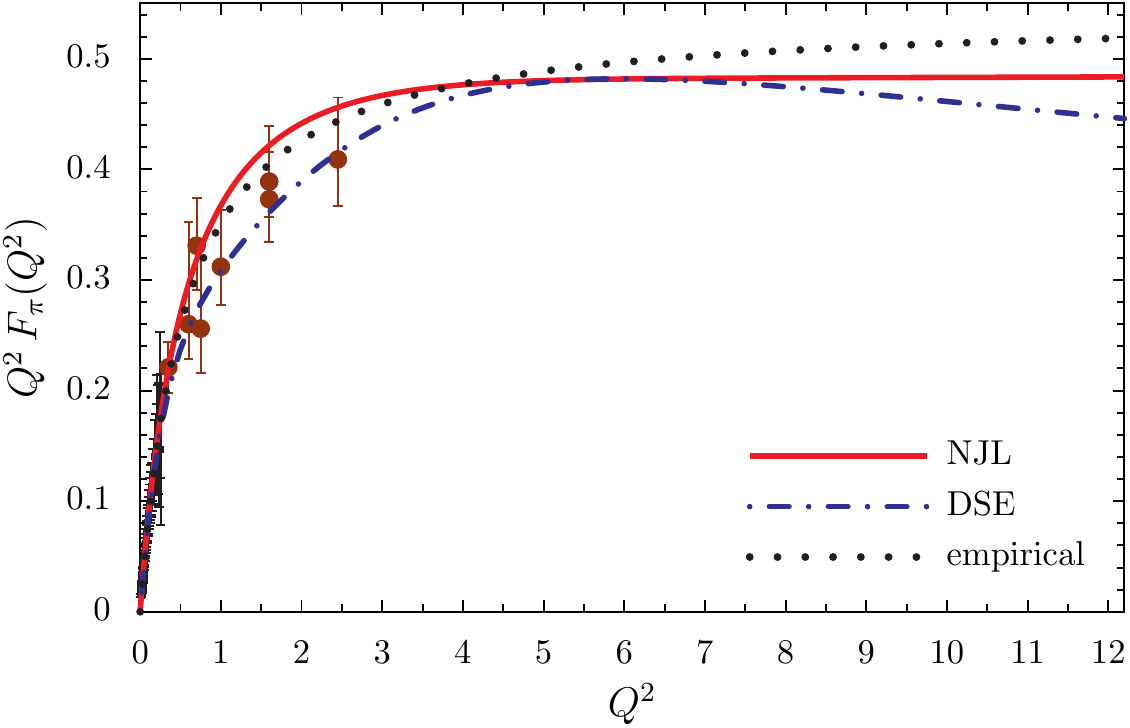}
\caption{(Colour online)
Results for $Q^2F_{\pi}(Q^2)$. See caption to Fig.~\ref{fig:pionff} for the nomenclature.}
\label{fig:Q2pionff}
\end{figure}

Results for the pion form factor  -- including effects from the 
dressed quark-photon vertex --  are presented in Figs.~\ref{fig:pionff} and \ref{fig:Q2pionff}, 
where comparisons to data~\cite{Amendolia:1984nz,Amendolia:1986wj,Horn:2006tm,Tadevosyan:2007yd,Huber:2008id,Blok:2008jy} , 
an empirical parametrization~\cite{Amendolia:1984nz} and the Dyson-Schwinger equation (DSE) 
result of Ref.~\cite{Chang:2013nia} have been made. We find excellent agreement with existing
data and the modest differences with the DSE result for $Q^2 \lesssim 6\,$GeV$^2$ are easily 
understood. The DSE result drops more rapidly that our NJL result primarily because the Bethe-Salpeter 
vertices in the DSE approach are non-pointlike and thereby suppress large relative moment 
between the dressed-quark and dressed-antiquark in the bound state. Our result for $Q^2\,F_\pi(Q^2)$
is very similar to the empirical monopole result and  begins to plateau for $Q^2 \gtrsim 6\,$GeV$^2$ 
where $Q^2\,F_\pi(Q^2) \simeq 0.49$. This maximum is almost identical to that obtained in the DSE, which is not surprising because in both
formalisms it is driven by dynamical chiral symmetry breaking~\cite{Klevansky:1992qe,Vogl:1991qt,Munczek:1994zz}. For 
$Q^2 \gtrsim 6\,$GeV$^2$ the DSE result for $Q^2\,F_\pi(Q^2)$ begins to decrease, 
which is a consequence of QCD's running coupling and a feature which is absent in our NJL
calculations.

%===============================================================================
%===============================================================================

%===============================================================================
%===============================================================================

\begin{figure}[tbp]
  \centering
  \begin{tikzpicture}
    \node[anchor=south west,inner sep=0] (image) at (0,0)    {\includegraphics[width=\columnwidth]{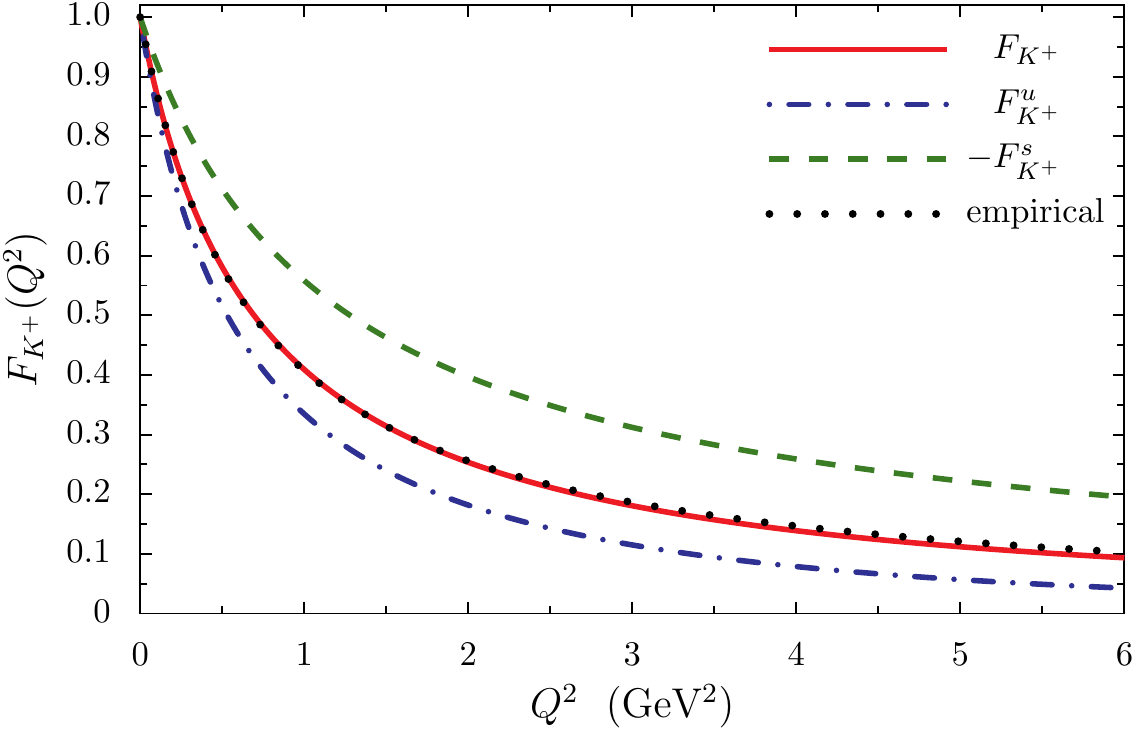}};
    \node[anchor=south west,inner sep=0] (image) at (1.8,3.93){\includegraphics[width=0.45\columnwidth]{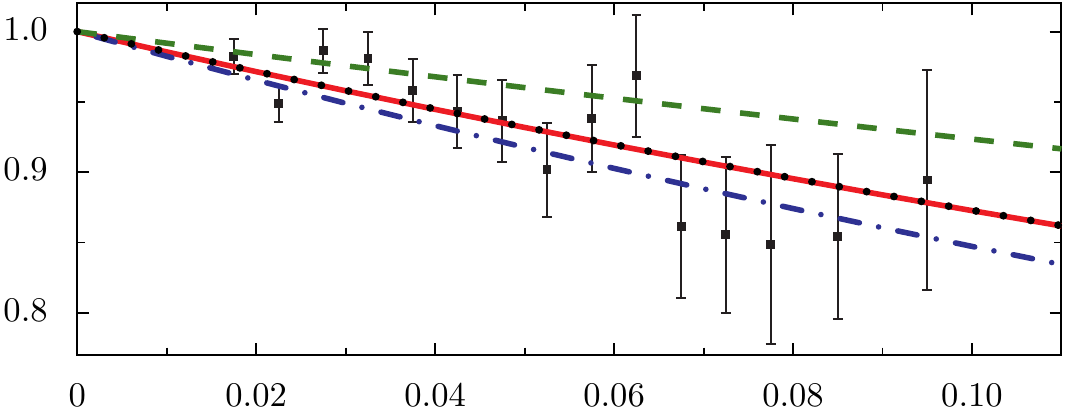}};
  \end{tikzpicture}
\caption{(Colour online) The $K^+$ form factor (solid line) together with the up 
(dashed-dotted line) and strange (dashed line) quark sector contributions. The dotted-line 
is the fit to data using the form $F_K(Q^2) = [1 + Q^2/\L_K^2]^{-1}$, giving $\L^2_K = 0.687\,$GeV$^2$, and
the insert compares our results with existing data taken from Ref.~\cite{Amendolia:1986ui}. }
\label{fig:kaonff}
\end{figure}

\begin{figure}[tbp]
\centering\includegraphics[width=\columnwidth]{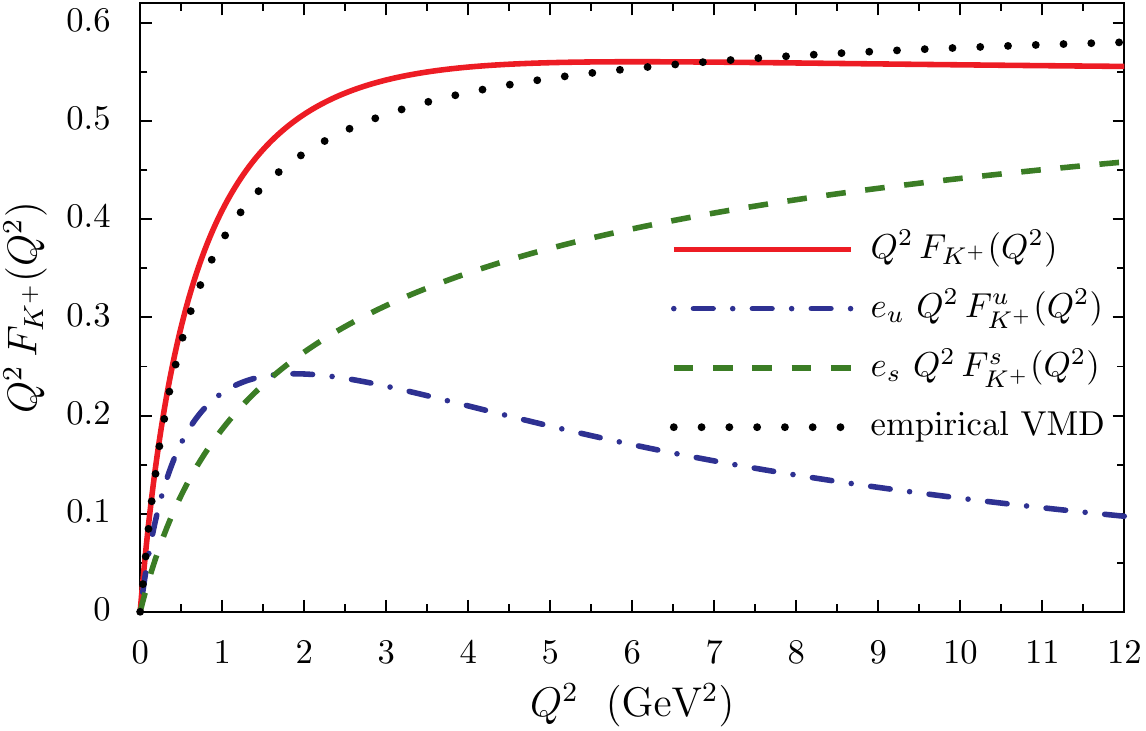}
\caption{(Colour online) Results for $Q^2\,F_{K^{+}} (Q^2)$ together with the charge-weighted 
quark-sector contributions and the empirical result obtained from Ref.~\cite{Amendolia:1986ui}.
This result clearly illustrates that the $s$ quark dominates the form factor at large $Q^2$.}
\label{fig:Q2kaonff} 
\end{figure}

\looseness=-1
Results for the $K^+$ form factor and the quark-sector components -- each including effects from the 
dressed quark-photon vertex -- are given in Figs.~\ref{fig:kaonff} and \ref{fig:Q2kaonff}. We find
excellent agreement with the data from Ref.~\cite{Amendolia:1986ui} and the empirical monopole 
$F_K(Q^2) = [1 + Q^2/\Lambda_K^2]^{-1}$ determined by reproducing the charge radius of  Ref.~\cite{Amendolia:1986ui}. 
In contrast to the pion, all existing data for the kaon form factor lies in the domain $0 < Q^2 < 0.1\,$GeV$^2$, 
and therefore we eagerly await any new data at $Q^2$ similar to the pion~\cite{Horn:2016rip}.
For the quark-sector form factors we observe a very large difference in their $Q^2$ evolution, with the $s$ quark
component much harder than the $u$ quark form factor. When weighted by the charges, as in Fig.~\ref{fig:Q2kaonff},
we find that the $s$ quark component begins to dominate the $K^+$ form factor for $Q^2 \geqslant 1.6\,$GeV$^2$, becoming 
completely dominant at very large $Q^2$.

%===============================================================================
\begin{table}[b]
\addtolength{\tabcolsep}{4.0pt}
\addtolength{\extrarowheight}{1.2pt}
\begin{tabular}{c|ccccccc}
\hline\hline
         & $r^{\text{exp't}}$           & $r$           & $r_u$  & $r_d$ & $r_s$     \\
\hline
$\pi^+$  & \ph{$-$}0.672 $\pm$ 0.008 & \ph{$-$}0.629 & 0.629  & $-$0.629       & 0      \\
$K^+$    & \ph{$-$}0.560 $\pm$ 0.031 & \ph{$-$}0.586 & 0.646  & 0              & $-$0.441  \\
$K^0$    &      $-$0.277 $\pm$ 0.018 &      $-$0.272 & 0      & \ph{$-$}0.646  & $-$0.441  \\
\hline\hline
\end{tabular}
\caption{Charge radius results for the pion and kaon, together with 
the various quark-sector contributions. All radii are in units of fm and the empirical results are from Refs.~\cite{Beringer:1900zz,Agashe:2014kda}.} 
\label{tab:radii}
\end{table}
%===============================================================================

%===============================================================================
%===============================================================================

%===============================================================================
%===============================================================================

Results for the pion and kaon radii are listed in Tab.~\ref{tab:radii}. For the pion 
we find a radius 6\% smaller than the Particle Data Group value~\cite{Agashe:2014kda} and 
agree within errors for both the $K^+$ and $K^0$ radii. We find that $r_{K^+}$ is about 
7\% smaller than $r_{\pi^+}$, which is driven by the quark-sector result $\lf|r^s_{K^+}\rg| < \lf|r^d_{\pi^+}\rg|$,
with $r^s_{K^+}/ r^d_{\pi^+} = 0.70$. We find the perhaps surprising result that $r^u_{K^+} > r^u_{\pi^+}$, with 
$r^u_K / r^u_\pi = 1.027$ a measure of environment sensitivity for the $u$ quark. These quark-sector radii 
are listed in Tab.~\ref{tab:radii}. As a measure of flavour 
breaking we have $\lf[r_{\pi^+} - r_{K^+}\rg]/\lf[r_{\pi^+} + r_{K^+}\rg] = 0.035$ and $[r^u_{K^+} + r^s_{K^+}]/[r^u_{K^+} - r^s_{K^+}] = 0.19$, which would vanish in the {\it SU}$(3)$ flavour limit. We therefore find that in some observables flavour breaking effects 
may be as large as 20\%.

%===============================================================================
\begin{figure}[tbp]
\centering\includegraphics[width=\columnwidth]{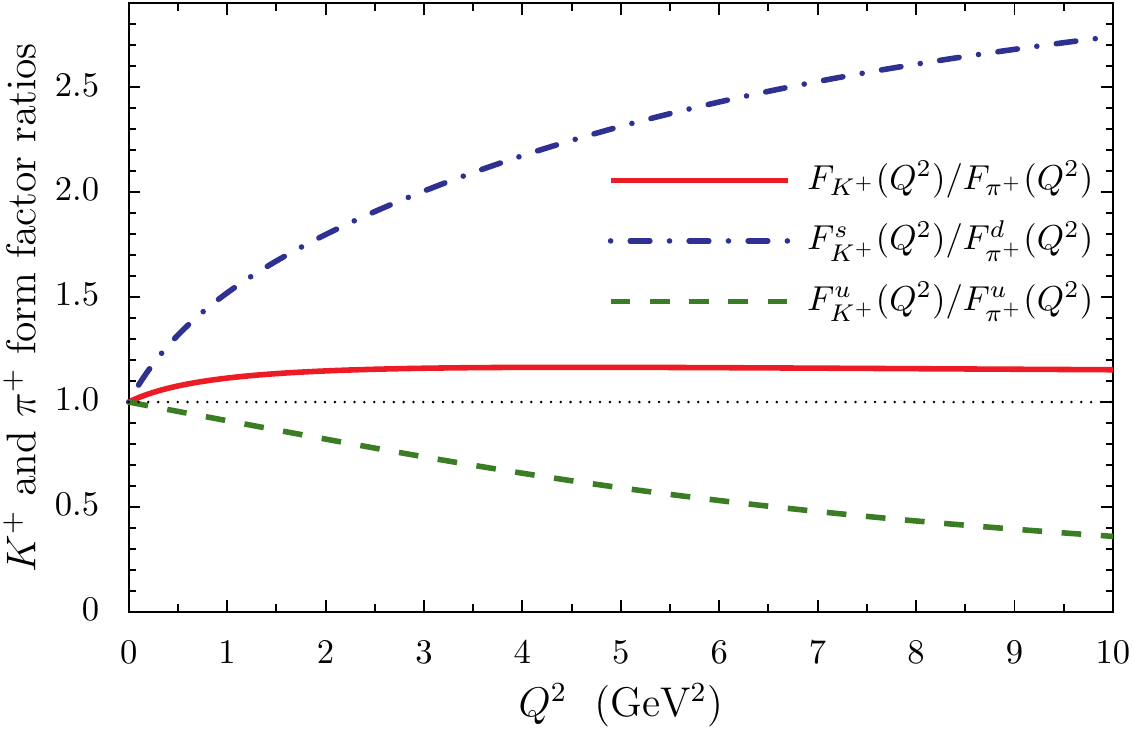}
\caption{(Colour online) We illustrate various pion and kaon form factor ratios, including
for the quark sector form factors, to ascertain a measure of flavour breaking and environment sensitivity
effects as a function of $Q^2$. Note, all ratios would be unity for all $Q^2$ in the $SU(3)$ flavour limit.}
\label{fig:ratios}
\end{figure}
%===============================================================================

In Fig.~\ref{fig:ratios} we illustrate the ratio $F_{K^+}(Q^2)/F_{\pi^+}(Q^2)$ which is
always greater than unity and becomes almost constant for $Q^2 \gtrsim 3\,$GeV$^2$. For 
very large $Q^2$ this ratio plateaus to the value $f_K^2/f_\pi^2 = 1.10$, 
in agreement with the QCD result in the conformal limit~\cite{Lepage:1979zb}:
\begin{align}
F_{K^+}(Q^2)/F_{\pi^+}(Q^2) \stackrel{Q^2 \gg \L_{\text{QCD}}}{\lra} f_K^2/f_\pi^2,
\end{align}
however we find $f_K = 97.3\,$MeV whereas the empirical value is $f_K = 110.4 \pm 0.8$~\cite{Agashe:2014kda}.
When expressed in terms of the quark sector form factors, and in the $m_u = m_d$ limit, we have
\begin{align}
\frac{F_{K^+}(Q^2)}{F_{\pi^+}(Q^2)} = e_u\,\frac{F^u_{K^+}(Q^2)}{F^u_{\pi^+}(Q^2)} - e_s\,\frac{F^s_{K^+}(Q^2)}{F^d_{\pi^+}(Q^2)},
\end{align}
where the various quark-sector ratios are also given in Fig.~\ref{fig:ratios}. It is clear 
therefore, that the large constant ratio $F_{K^+}(Q^2) / F_{\pi^+}(Q^2)$ conceals dramatic
flavour breaking effects in the quark-sector form factors that grow with increasing $Q^2$.
In the {\it SU(3)} flavour limit all ratios in Fig.~\ref{fig:ratios} would be unity for all $Q^2$.
However, at $Q^2 = 10\,$GeV$^2$ we find $F^u_{K^+} / F^u_{\pi^+} \simeq 0.36$ and 
$F^s_{K^+} / F^d_{\pi^+} \simeq 2.74$. Therefore, at large $Q^2$ we find very large flavour breaking
and environment sensitivity effects. The final ratio illustrated in Fig.~\ref{fig:ratios} is
$F^u_{K^+}(Q^2) / F^s_{\pi^+}(Q^2)$, which rapidly drops to zero with increasing $Q^2$. This
behaviour can be understood by noting that a form factor is a measure of the ability of a hadron
to absorb an electromagnetic current and remain a hadron. In the case of the $K^+$, if the $u$-quark 
interacts with the electromagnetic current it must drag along the heavier $s$-quark for the $K^+$
to remain intact, which becomes increasingly more difficult at larger $Q^2$ than if the struck 
quark is an $s$-quark. Therefore this ratio may well be a very sensitive measure of confinement 
effects in QCD.

%===============================================================================
\begin{figure}[tbp]
\centering\includegraphics[width=1.0\columnwidth]{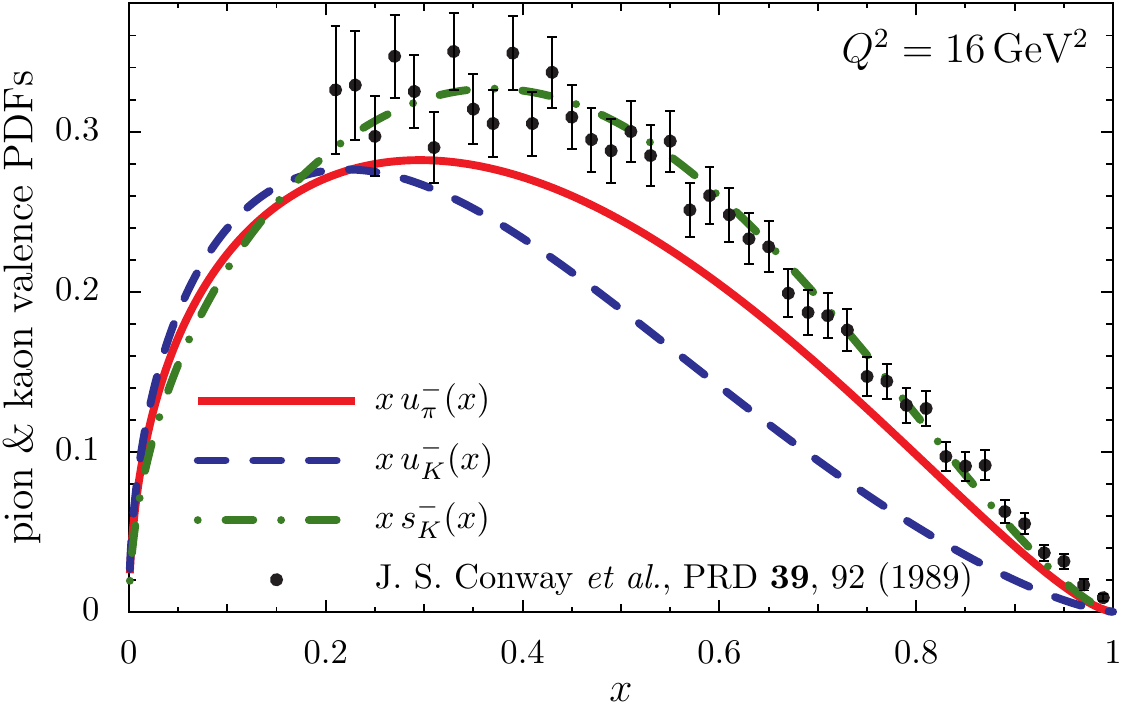}
\caption{(Colour online) Results for the valence quark distributions 
of the $\pi^{+}$ and $K^{+}$, evolved from the model scale using the NLO DGLAP
equations~\cite{Gribov:1972rt,Altarelli:1977zs,Dokshitzer:1977sg,Miyama:1995bd}. The solid line represents the valence $u$ or $\bar{d}$ PDF
in the $\pi^{+}$, the dot-dashed line is the valence $\bar{s}$ quark and the 
dashed line the valence $u$ quark in the $K^{+}$. The experimental data 
are taken from Ref.~\cite{Conway:1989fs}.}
\label{fig:pdfs} 
\end{figure}
%===============================================================================

%===============================================================================
%===============================================================================
\section{PARTON DISTRIBUTION FUNCTION RESULTS}
Results for the pion and kaon valence PDFs at $Q^2 = 16 \,$GeV$^2$ are presented in 
Fig.~\ref{fig:pdfs} and compared to empirical data for the pion valence PDF from 
Ref~\cite{Conway:1989fs}.\footnote{This data has been reanalyzed in Ref.~\cite{Sutton:1991ay}, where the new
empirical parametrization would imply that the data shown in Fig.~\ref{fig:pdfs} should be shifted 
down for large-$x$ and shifted up for moderate-$x$.}
We find reasonable agreement over the entire $x$ domain 
where data is available. Our results have been evolved using the next-to-leading order (NLO) DGLAP evolution
equations~\cite{Gribov:1972rt,Altarelli:1977zs,Dokshitzer:1977sg,Miyama:1995bd} from a model scale of $Q_0^2 = 0.16\,$GeV$^2$, which was independently 
determined in Ref.~\cite{Cloet:2005pp} in the study of nucleon PDFs. At the model scale we find that 
the momentum fraction carried by the $u$ and $s$ quarks in the $K^+$ equal $\lf<x\,u\rg> = 0.42$ and 
$\lf<x\,s\rg> = 0.58$ (at this scale gluons carry no momentum so these results saturate 
the momentum sum rule). We therefore find flavour breaking effects of 
$\lf[\lf<x\,s\rg> - \lf<x\,u\rg>\rg] / \lf[\lf<x\,s\rg> + \lf<x\,u\rg>\rg] \simeq 16$\% which 
is similar to that seen in the masses: $\lf[M_s - M_u\rg] / \lf[M_s + M_u\rg] \simeq 21$\% and
quark-sector radii.
As another measure of {\it SU(3)} flavour breaking we note that at the model scale $u_K(x)$
peaks at $x_u = 0.237$ and  $\bar{s}_K(x)$ peaks at $x_s = 1 - x_u = 0.763$, which
implies flavour breaking effects of around $\lf[x_s - x_u\rg] / \lf[x_s + x_u\rg] \simeq 53$\%.
Note that in the {\it SU(3)} flavour limit these distributions would peak at $x = 0.5$, which is
the case for the pion when $m_u = m_d$.

\begin{figure}[tbp] 
\centering\includegraphics[width=1.0\columnwidth]{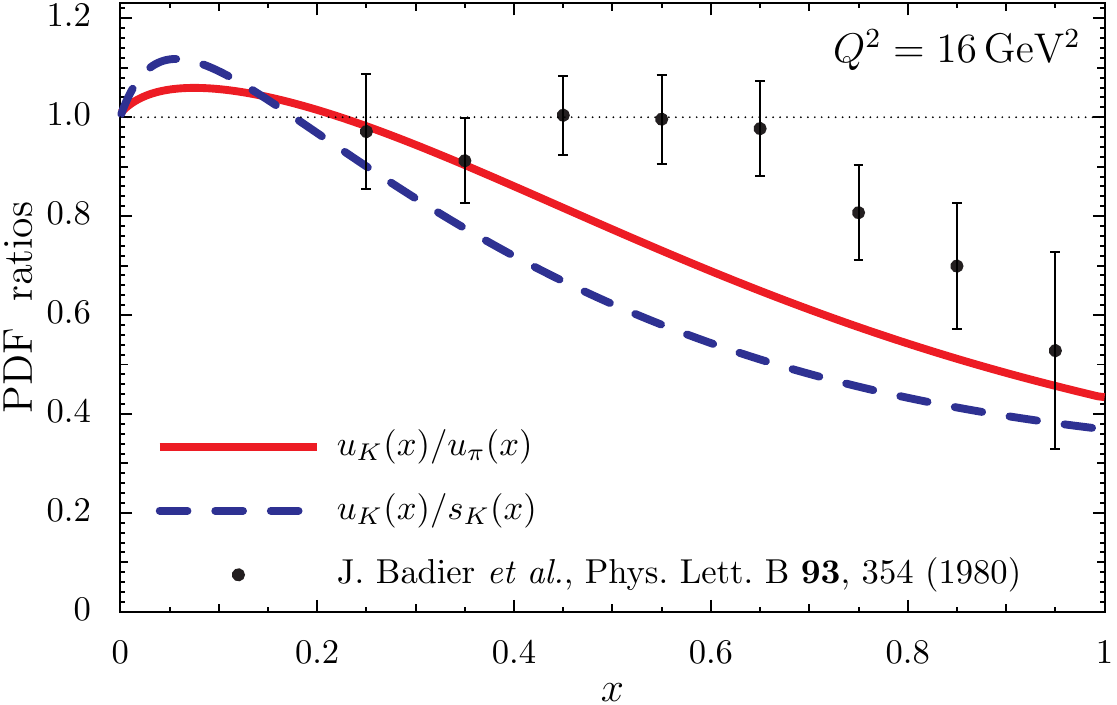}
\caption{(Colour online) The solid line gives the ratio of the $u$ quark distribution of the kaon to the $u$ quark distribution of the pion, after NLO evolution to $Q^2$ = 16 $\rm{GeV}^2$. The dashed line gives the ratio of the $u$ quark to $s$ quark distributions in the kaon at $Q^2$ = 16 $\rm{GeV}^2$.}
\label{fig:pdfratio} 
\end{figure}

The ratio $u_{K^+}(x)/u_{\pi^+}(x)$ is illustrated in Fig.~\ref{fig:pdfratio} 
at $Q^2 = 16\,$GeV$^2$, however this ratio has only a slight $Q^2$ dependence and
in the limit $x \to 1$ is a fixed point in $Q^2$. We find $u_{K^+}/u_{\pi^+} \to 0.434 \simeq M_u^2/M_s^2$ as $x \to 1$,
in good agreement with existing data from Ref.~\cite{Badier:1980jq}. However, the $x$ dependence differs from much of the data 
in the valence region, the reason for this discrepancy is not clear, however it may
lie with the absence of momentum dependence in standard NJL Bethe-Salpeter vertices~\cite{Nguyen:2011jy,Maris:2003vk}, or 
with the data itself.
We note however the correspondence that $u_{K^+}/u_{\pi^+} < 1$ as $x \to 1$ and that 
$F^u_K(Q^2)/F_\pi^u(Q^2) < 1$ for $Q^2 \gg \L_{\text{QCD}}^2$. 
Fig.~\ref{fig:pdfratio} also illustrates the ratio $u_{K^+}(x)/s_{K^+}(x)$, which 
approaches $0.37$ as $x \to 1$. It is evident that flavour breaking effects have 
a sizable $x$ dependence, being maximal at large $x$ while becoming negligible 
at small $x$ where perturbative effects from DGLAP evolution dominate.

%===============================================================================
%============================================================================   

The limit $x \to 1$ corresponds to elastic scattering from the target 
and as such it is natural to expect a correspondence between form factors
and PDFs in this limit. Such a correspondence was first considered by
Drell and Yan~\cite{Drell:1969km}, and West~\cite{West:1970av}, finding the relation:
\begin{align}
F(Q^2) \stackrel{Q^2 \gg \L_{\text{QCD}}}{\sim} \frac{1}{Q^{2n}}  \quad \Longleftrightarrow \quad q(x) \stackrel{x\to 1}{\sim} (1-x)^{2n-1},
\end{align}
between a hadron's form factor and PDF, where $n$ is the number of spectators. For the pion the expectation is 
$F_\pi(Q^2) \stackrel{Q^2 \gg \L_{\text{QCD}}}{\sim} 1/Q^2$ and therefore 
the Drell-Yan-West (DYW) relation implies 
$q_\pi(x) \stackrel{x\to 1}{\sim} (1-x)$, in good agreement with the data 
in Fig.~\ref{fig:pdfs}. For the pion however the DYW relation is in 
disagreement with the more rigorous QCD analyses of Refs.~\cite{Farrar:1979aw,Lepage:1979zb}, that find
\begin{align}
F_\pi(Q^2) \stackrel{Q^2 \gg \L_{\text{QCD}}}{\sim} \frac{1}{Q^{2}}  
\quad \Longleftrightarrow \quad q_\pi(x) \stackrel{x\to 1}{\sim} (1-x)^{2}.
\end{align}
The conclusion argued therefore in Ref.~\cite{Lepage:1979zb} is the the DYW relation
is not generally valid in QCD, although it does appear to hold for baryon
states. At the model scale our NJL calculation for the pion satisfies
\begin{align}
F_\pi(Q^2) \stackrel{Q^2 \gg \L_{\text{QCD}}}{\sim} \frac{1}{Q^{2}}  
\quad \Longleftrightarrow \quad q_\pi(x) \stackrel{x\to 1}{\sim} (1-x)^0.
\end{align}
and therefore does not agree with the DYW relation. On the other hand, after
DGLAP evolution to $Q^2 \sim 10\,$GeV$^2$ the pion and kaon PDFs do behave as
$q_\pi(x) \stackrel{x\to 1}{\sim} (1-x)^1$ as evident in Fig.~\ref{fig:pdfs}. 
As a reflection of the expectations of what may be expected by DYW-like 
relations we find that $u_{K^+} / s_{K^+} < 1$ as $x\to 1$ and $\lf|F^u_{K^+}/F_{K^+}^s\rg| < 1$ for
$Q^2 \gg \L_{\text{QCD}}$.

%===============================================================================
%===============================================================================
\section{SUMMARY \label{conclusion}}
We have used the NJL model -- with proper-time regularization 
to simulate the effect of confinement -- to calculate the 
electromagnetic form factors and PDFs of the pion and kaon. 
For the former we included the effect of vertex dressing 
through vector meson like correlations in the $t$-channel, which 
do not contribute to the PDFs. Particular attention was paid to the 
individual quark flavour contributions and the associate flavour breaking
and environment sensitivity effects.

This work produced several remarkable results. Firstly, as illustrated 
in Figs.~\ref{fig:kaonff}--\ref{fig:ratios}, 
the effect of the larger mass of the strange quark 
on the electromagnetic form factors is dramatic. Indeed, even 
though $\lf|e_s\rg| < \lf|e_u\rg|$ the $s$-quark dominates 
the total elastic form factor of the $K^+$ for large $Q^2$. 
Surprisingly, as shown in Fig.~\ref{fig:ratios}, 
even though there are very significant 
changes in the individual flavour contributions in the kaon, 
the total pion and kaon form factors lie within about 10-15\%
for all $Q^2$, with the environmental suppression of the 
$u$-quark from factor in the $K^+$ more or less compensated
by the increase in the strange quark from factor over that 
of the $d$ quark. In terms of the overall agreement with experiment, 
the total kaon form factor agrees very well with the limited 
existing data. In the case of the pion, the data extends to 
much larger $Q^2$, where again we find excellent agreement. 

The effects of the strange quark mass on the PDFs is less 
spectacular. In Fig.~\ref{fig:pdfs} 
we saw that the strange quark PDF in the $K^+$ is considerably enhanced 
over that of the $u$-quark in the valence region.
%In addition, $u_{\pi^+}$ does agree with the data of Conway and collaborators. 
Most importantly, as we see in Fig.~\ref{fig:pdfratio}, 
the empirical suppression of $u_{K^+}$ compared 
with $u_{\pi^+}$ is rather well described. 

\looseness=-1
The comparison of the asymptotic behaviour of the individual flavour 
form factors and parton distributions is fascinating. While all 
elastic form factors in this model behave as $1/Q^2$ at larger $Q^2$, 
$F^{s}_{K}(Q^2)/F^u_K(Q^2) \sim 10$ at $Q^2 = 10\,$GeV$^2$. Nevertheless, as already noted, 
the total $K^+$ and $\pi^+$ form factors only differ by 10-15\%. Numerous other effects of
flavour breaking have also been determined, for example, the pion and kaon charge radii, 
where effects of around 20\% were typically observed.

\begin{acknowledgments}
This work was supported by the U.S. Department of Energy, Office of Science, 
Office of Nuclear Physics, contract no. DE-AC02-06CH11357 and the Australian 
Research Council through the ARC Centre of Excellence in Particle Physics 
at the Terascale and an ARC Australian Laureate Fellowship FL0992247 at
the University of Adelaide.
\end{acknowledgments}

%\bibliographystyle{/home/icloet/.files/myapsrev4-1}
%\bibliography{bibtexfile,bibtexfile_cloet,bibtexfile_arXiv,bibtexfile_misc,bibtexfile_books}

\end{document}